\documentclass[reqno,a4paper,final,11pt]{amsart}
\usepackage[utf8]{inputenc}
\usepackage{amssymb,amstext,amsthm,amscd,mathrsfs}
\usepackage{color}
\usepackage{array}
\usepackage{hyperref}
\hypersetup{%
  pdftitle   = {The Killing superalgebra of ten-dimensional
    supergravity backgrounds},
  pdfkeywords = {supergravity, heterotic, type II, type I, heterotic,
    supersymmetry, superalgebra, Killing spinors, homogeneity},
  pdfauthor  = {José Figueroa-O'Farrill, Emily Hackett-Jones, George Moutsopoulos},
  pdfcreator = {\LaTeX\ with package \flqq hyperref\frqq}
}
\newcommand{\half}{\tfrac12}

\newcommand{\bx}{\boldsymbol{x}}

\newcommand{\be}{\boldsymbol{e}}
\newcommand{\beps}{\boldsymbol{\varepsilon}}
\newcommand{\fg}{\mathfrak{g}}

\newcommand{\Cl}{\mathrm{C}\ell}
\newcommand{\fspin}{\mathfrak{spin}}
\newcommand{\fsp}{\mathfrak{sp}}
\newcommand{\fsl}{\mathfrak{sl}}
\newcommand{\fgl}{\mathfrak{gl}}
\newcommand{\fsu}{\mathfrak{su}}
\newcommand{\fu}{\mathfrak{u}}
\newcommand{\Spin}{\mathrm{Spin}}
\ifx\Sp\UnDeFiNeD
\newcommand{\Sp}{\mathrm{Sp}}
\else
\renewcommand{\Sp}{\mathrm{Sp}}
\fi

\newcommand{\SL}{\mathrm{SL}}
\newcommand{\GL}{\mathrm{GL}}
\newcommand{\SO}{\mathrm{SO}}
\renewcommand{\O}{\mathrm{O}}

\newcommand{\RR}{\mathbb{R}}
\newcommand{\GG}{\mathbb{G}}
\renewcommand{\SS}{\mathbb{S}}

\newcommand{\eL}{\mathscr{L}}

\DeclareMathOperator{\End}{End}
\DeclareMathOperator{\Mat}{Mat}

\DeclareMathOperator{\Symm}{Sym}

\DeclareMathOperator{\rank}{rank}
\newcommand{\MUNCH}[1]{\relax}

\allowdisplaybreaks[1]
\begin{document}
\title[Killing superalgebra of ten-dimensional backgrounds]{The
  Killing superalgebra of ten-dimensional supergravity backgrounds}
\author[Figueroa-O'Farrill]{José Figueroa-O'Farrill}
\author[Hackett-Jones]{Emily Hackett-Jones}
\author[Moutsopoulos]{George Moutsopoulos}
\address{Maxwell Institute and School of Mathematics, University of
  Edinburgh, UK}
\begin{abstract}
  We construct the Killing superalgebra of supersymmetric backgrounds
  of ten-dimensional heterotic and type II supergravities and prove
  that it is a Lie superalgebra.  We also show that if the fraction
  of supersymmetry preserved by the background is greater than $1/2$,
  in the heterotic case, or greater than $3/4$ in the type II case,
  then the background is locally homogeneous.
\end{abstract}
\maketitle
\tableofcontents

\section{Introduction}
\label{sec:intro}

One of the major problems in supergravity is to understand how
supersymmetry shapes the geometry of classical solutions to the field
equations.  In other words, what is the geometry of a supergravity
background which preserves a given fraction of the supersymmetry?
Much recent work has gone into answering this question for
supergravity theories in diverse dimensions.  For some of the
supergravity theories in dimensions $4$, $5$ and $6$, there is a more
or less complete story \cite{Tod1,Tod2,GGHPR,GMR,CFOSchiral}, whereas
for the ten- and eleven-dimensional supergravities the story is far
from complete.  At the high end of the supersymmetry fraction, there
is a complete classification of maximally supersymmetric backgrounds
\cite{FOPMax} as well as non-existence results
\cite{NoIIBPreons,PreonsConfinement,NoMPreons,FigGadPreons} for
so-called preonic backgrounds \cite{11dPreons,Preons} preserving a
fraction $\nu=\tfrac{31}{32}$.  At the low end of the fraction, e.g.,
$\nu=\frac1{32}$, local expressions for the metric and fluxes have
been derived using the various interplays between Killing spinors and
differential forms: either by considering the reduction of the
structure group of the manifold brought about by the existence of
differential forms built out of Killing spinors, the so-called
``$G$-structure'' approach of \cite{GauPak,GauGutPak}, or else by
thinking of spinors themselves as differential forms, in the so-called
``spinorial geometry'' approach of
\cite{GGP1,GGP2,GGPR,GLP,GPRSTypeI}.

A somewhat less ambitious aim is to find general properties of
backgrounds preserving less than maximal but more than minimal
fractions of supersymmetry.  One such property is local homogeneity.
It was shown in \cite{FMPHom} that M-theory backgrounds preserving a
fraction $\nu > \frac34$ of the supersymmetry are locally homogeneous;
that is, at every point there is a frame for the tangent space
consisting of flux-preserving Killing vectors.  This was proven by
first constructing a Lie superalgebra (termed here the \emph{Killing
  superalgebra} due to the nature of its construction, to be reviewed
below) associated to a supersymmetric M-theory background, along the
lines traced in \cite{AFHS, GMT1, GMT2, PKT, JMFKilling, FOPFlux,
  NewIIB, ALOKilling} and then investigating the dimension of the
translational component of the image of the squaring map from Killing
spinors to Killing vectors.

In this paper we address this issue for the ten-dimensional
supergravity theories, both type I (coupled to supersymmetric
Yang--Mills) and type II.  The paper is organised as follows.  In
§\ref{sec:typeI} we discuss type I supergravity coupled to
supersymmetric Yang--Mills: the theory and the Killing spinors are
discussed in §\ref{sec:Itheory}, whereas the Killing superalgebra and
the homogeneity of sufficiently supersymmetric backgrounds is
discussed in §\ref{sec:IKSA} and §\ref{sec:typeIHom}, respectively.
In particular we show that the $\nu_c= \half$ is the critical fraction
above which local homogeneity is guaranteed.  The more involved case
of type IIB supergravity is discussed in §\ref{sec:typeIIB}.  In
§\ref{sec:IIBtheory} we discuss the theory and Killing spinors,
respectively.  The Killing superalgebra is constructed in
§\ref{sec:IIBKSA}, where we prove that this is indeed a Lie
superalgebra.  This requires checking that the Jacobi identities are
satisfied.  The only nontrivial identity is the odd-odd-odd identity,
which is shown to hold in §\ref{sec:IIBjac} making use of Fierz
identities proven in Appendix~\ref{app:fierz}.  The question of local
homogeneity is addressed in §\ref{sec:IIBHom}, where we find, just as
with M-theory backgrounds, that the critical fraction above which
local homogeneity is guaranteed satisfies $\half \leq \nu_c \leq
\frac34$.  Finally, in §\ref{sec:conclusion} we offer some concluding
remarks and in particular show how to obtain results for backgrounds
of type I and IIA supergravities from the known results for heterotic
and eleven-dimensional supergravities.  The paper contains three
appendices.  In the first, Appendix~\ref{app:spinors}, we set out our
spinorial conventions and notations.  The reader is urged to read this
appendix first, at least to skim the notation.  The second,
Appendix~\ref{app:KG3G5}, contains some of the calculations used in
§\ref{sec:closure}.  Finally Appendix~\ref{app:fierz} collects the
Fierz identities mentioned above.

\section{Heterotic supergravity backgrounds}
\label{sec:typeI}

The field theory limit of the heterotic string \cite{HeteroticString}
is ten-dimensional $N{=}1$ supergravity \cite{ChamseddineN=1} coupled
to $N{=}1$ supersymmetric Yang--Mills \cite{superYM}.  This theory was
constructed in \cite{ChaplineManton} generalising the construction in
\cite{10dsugraMaxwell} for the abelian theory.

\subsection{Supersymmetric heterotic backgrounds}
\label{sec:Itheory}

The bosonic fields in this theory are a ten-dimensional lorentzian
metric $g$, the dilaton $\phi$, the NS-NS $3$-form $H$ and a gauge
field-strength $F$.  In heterotic string theory the gauge group is
constrained, but we will work with arbitrary gauge group in what
follows.  The $3$-form $H$ is assumed to be closed in the supergravity
limit; although this equation receives $\alpha'$ corrections.

The equations of motion are obtained by varying the following action
(in the string frame)
\begin{equation}
  \label{eq:action}
    I = \int d^{10}x \sqrt{-g} e^{-2\phi} \left( R + 4 |d\phi|^2 - \half
    |H|^2 - \tfrac{1}{2} |F|^2 \right)~.
\end{equation}
The supersymmetry parameters are chiral spinors: they are real and
have sixteen components.  Killing spinors satisfy three equations, a
differential equation coming from the variation of the gravitino, and
two algebraic equations coming from varying the dilatino and the
gaugino, respectively:
\begin{equation}
  \label{eq:typeIKS}
  \begin{aligned}[m]
    D_m \varepsilon &:= \left(\nabla_m - \tfrac18
      H_{mnp}\gamma^{np}\right) \varepsilon = 0\\
    P \varepsilon &:= \left(\partial_m\phi \gamma^m - \tfrac1{12}
      H_{mnp}\gamma^{mnp}\right) \varepsilon = 0\\
    Q \varepsilon &:= \half F_{mn}\gamma^{mn} \varepsilon = 0~.
  \end{aligned}
\end{equation}
The connection $D$ is metric compatible but has totally antisymmetric
torsion, given by $H$.  As these equations are linear, the set of
Killing spinors is a real vector space, denoted $\fg_1$ and of
dimension $16\nu$, where $0\leq \nu \leq 1$ is the fraction of
supersymmetry preserved by the background $(M,g,\phi,H,F)$.  A lot of
progress has been made on the classification of supersymmetric
heterotic backgrounds \cite{FOPMax,FKYHeterotic,GLP}.

A Killing vector which preserves not just the metric but also $\phi$
and $H$, and which preserves $F$ up to a gauge transformation, is said
to be an infinitesimal symmetry of the background.  It is clear that
they too form a vector space $\fg_0$.  More is true, however, and as
they close under the Lie bracket of vector fields, they form a Lie
algebra.  In the next section we will show that just as in the case of
M-theory backgrounds, the vector superspace $\fg = \fg_0 \oplus \fg_1$
carries the structure of a Lie superalgebra extending the Lie algebra
structure of $\fg_0$.

\subsection{The Killing superalgebra}
\label{sec:IKSA}

In this section we construct a Lie superalgebra structure on the
superspace $\fg := \fg_0 \oplus \fg_1$.  The construction is
well-known by now and has been explained in detail before
\cite{JMFKilling,FMPHom}.  There are three components to the Lie
bracket: the Lie bracket on $\fg_0$, given by restricting the Lie
bracket of vector fields to the infinitesimal symmetries; the action
of $\fg_0$ on $\fg_1$ given by the spinorial Lie derivative
\cite{Kosmann}, and the squaring of spinors, which gives rise to the
bracket $\fg_1 \otimes \fg_1 \to \fg_0$.

We begin by showing that the vector field obtained by squaring a
Killing spinor is an infinitesimal symmetry.

The spin-invariant inner product on spinors has the following symmetry
properties:
\begin{equation*}
  \overline\varepsilon_1 \gamma^{n_1\cdots n_p}\varepsilon_2 =
  - (-1)^{p(p+1)/2} \overline\varepsilon_2 \gamma^{n_1\cdots
    n_p}\varepsilon_1~,
\end{equation*}
whence it is symmetric for $p\equiv 1,2 \pmod 4$ and antisymmetric
otherwise; although if $\varepsilon_i$ are chiral spinors of the same
chirality, this will vanish unless $p$ is odd.

Let $\varepsilon$ be a Killing spinor and let $K^m =
\overline\varepsilon \gamma^m \varepsilon$ denote its Dirac current.
Since $D_m \varepsilon = 0$ and $D$ is a spin connection, it follows
trivially that $D_m K^n = 0$.  Since $D$ is metric-compatible, we can
lower the index to obtain
\begin{equation*}
  D_m K_n = \nabla_m K_n - \half H_{mnp} K^p = 0~.
\end{equation*}
Symmetrizing, we find that $\nabla_mK_n + \nabla_n K_m = 0$, so that
$K$ is a Killing vector, and antisymmetrizing we find that the
contraction of $H$ by $K$ is exact:
\begin{equation}
  \label{eq:exact}
  K^p H_{pmn} = (dK)_{mn}~.
\end{equation}
In turn this shows that, since $H$ is closed, $\eL_K H = d\imath_K H =
0$.  Hence $H$ is also left invariant by $K$.

To show that $\phi$ is also left invariant, we take the inner product
of the dilatino variation $P\varepsilon = 0$ with $\varepsilon$.
There are two terms: a symmetric term with one gamma matrix and an
antisymmetric term with three.  This latter term vanishes by symmetry,
whence we remain with 
\begin{equation*}
  0 = \overline\varepsilon P \varepsilon = \partial_m\phi \overline\varepsilon
  \gamma^m \varepsilon = K^m \partial_m \phi~,
\end{equation*}
which shows that $K$ leaves $\phi$ invariant.

Finally we show that $K$ leaves $F$ invariant up to a gauge
transformation.  To do this we take the inner product of the gaugino
equation with $\varepsilon$ after multiplying it with a gamma matrix,
to obtain
\begin{equation*}
  0 = \overline\varepsilon \gamma_n Q \varepsilon =  F_{mn}
  \overline\varepsilon \gamma^n \varepsilon~,
\end{equation*}
after using that $\overline\varepsilon \gamma^{mnp} \varepsilon = 0$
by symmetry.  The above equation says that the contraction of $F$ by
$K$ vanishes, which implies that $F$ is invariant up to gauge
transformations.  Indeed,
\begin{equation*}
  \eL_K F = d\imath_K F + \imath_K dF = \imath_K dF~.
\end{equation*}
Using the Bianchi identity $dF = - [A,F]$, where $A$ is the
corresponding gauge field, and again the fact that $\imath_K F = 0$, we
see \begin{equation*}
  \eL_K F = - [\imath_K A, F]~,
\end{equation*}
which is an infinitesimal gauge transformation with parameter
$-\imath_K A$.  It is always possible to choose the ``temporal'' gauge
$\imath_K A = 0$, in which $F$ is truly invariant under $K$.

In summary, we have shown that $K$ is an infinitesimal symmetry of the
background.  Let us remark that by the standard polarization
trick, the vector $\overline\varepsilon_1 \gamma^m\varepsilon_2$, for
$\varepsilon_1$ and $\varepsilon_2$ Killing spinors, is also an
infinitesimal symmetry.

We now verify that the natural action of $\fg_0$ on $\fg_1$ actually
preserves the space of Killing spinors.  As explained, for instance,
in \cite{FMPHom}, the bracket $\fg_0 \otimes \fg_1 \to \fg_1$ is given
by the spinorial Lie derivative \cite{Kosmann}
\begin{equation*}
  \eL_K \varepsilon = K^m \nabla_m \varepsilon + \tfrac18
  (dK)_{mn}\gamma^{mn} \varepsilon~,
\end{equation*}
where the sign in the second term is due to the sign in the Clifford
algebra.  For $\varepsilon$ a Killing spinor,
\begin{equation*}
  \nabla_m \varepsilon = \tfrac18 H_{mnp}\gamma^{np}\varepsilon~,
\end{equation*}
whence using \eqref{eq:exact}
\begin{equation}
  \label{eq:spinLie}
  \eL_K \varepsilon = \tfrac14 K^m H_{mnp}\gamma^{np}\varepsilon~.
\end{equation}

We will now show that if $\varepsilon$ is a Killing spinor, so is
$\eL_K \varepsilon$ for all $K \in \fg_0$.  A standard property of the
spinorial Lie derivative is that for any Killing vector $K$, 
\begin{equation}
  \label{eq:Lnabla}
  \eL_K \nabla_X \varepsilon = \nabla_X \eL_K \varepsilon +
  \nabla_{[K,X]} \varepsilon~,
\end{equation}
and the same is true for $D$ in place of $\nabla$ provided that $K$
also preserves $H$.  As a consequence of this identity, the Lie
derivative along $K \in \fg_0$ of a $D$-parallel spinor is again
$D$-parallel.  Furthermore, since $K$ also preserves $\phi$ and $F$
(up to gauge transformations), the Lie derivative along $K$ of a
spinor satisfying the dilatino and gaugino equations---the second and
third equations in \eqref{eq:typeIKS}, respectively---will again
satisfy these equations.  In summary, the spinorial Lie derivative
defines a linear map $\fg_0 \otimes \fg_1 \to \fg_1$.

We now have defined each of the graded pices of the bracket $\fg
\otimes \fg \to \fg$ and it remains to show that the Jacobi identity
is satisfied.  This identity breaks up into types:
$(\fg_0,\fg_0,\fg_0)$, $(\fg_0,\fg_0,\fg_1)$, $(\fg_0,\fg_1,\fg_1)$
and $(\fg_1,\fg_1,\fg_1)$.  All but the last are true for reasons
which are more or less geometrically obvious: properties of the Lie
bracket of vector fields and of the spinorial Lie derivative, as
discussed in \cite{JMFKilling,FMPHom}.  The $(\fg_1,\fg_1,\fg_1)$
Jacobi identity usually has to be checked and this case is no
exception.  This Jacobi identity is equivalent to
$[\varepsilon,[\varepsilon,\varepsilon]]=0$ for all $\varepsilon\in
\fg_1$.  In our geometric realisation, this is equivalent to
\begin{equation*}
  \eL_K \varepsilon = 0\qquad\text{for all Killing spinors
    $\varepsilon$,}
\end{equation*}
where $K^m = \overline\varepsilon \gamma^m \varepsilon$ is Dirac
current of $\varepsilon$.  Since, as we have seen above, $DK=0$, it
follows from equation \eqref{eq:spinLie}, that all we need to show is
that
\begin{equation*}
  K^m H_{mnp} \gamma^{np} \varepsilon = 0\qquad\text{for all Killing
    spinors $\varepsilon$.}
\end{equation*}
This identity can be proven by the following simple argument.  Let
us rewrite the left-hand side as
\begin{equation*}
  \tfrac16 K^\ell H_{mnp} \left(\gamma_\ell \gamma^{mnp} +
    \gamma^{mnp} \gamma_\ell\right) \varepsilon~,
\end{equation*}
which, using the dilatino variation $P\varepsilon = 0$, can be
rewritten further as
\begin{equation*}
  2 K^\ell \partial_m \phi \gamma_\ell \gamma^m \varepsilon +
  \tfrac16 K^\ell H_{mnp} \gamma^{mnp} \gamma_\ell \varepsilon~.
\end{equation*}
Since $K^m\partial_m\phi = 0$, we can further rewrite this as
\begin{equation*}
  -2 K^\ell \partial_m \phi \gamma^m \gamma_\ell \varepsilon +
  \tfrac16 K^\ell H_{mnp} \gamma^{mnp} \gamma_\ell \varepsilon~.
\end{equation*}
which vanishes because $K^\ell \gamma_\ell \varepsilon = 0$.  Indeed,
since both $\varepsilon$ and $K$ are covariantly constant with respect
to the connection $D$, it is sufficient to prove this at a point,
where it can be seen to follow from representation theory alone, as
explained in Appendix~\ref{app:spinors}.

\subsection{Homogeneity of heterotic backgrounds}
\label{sec:typeIHom}

In \cite{FMPHom} it is shown that M-theory backgrounds with enough
supersymmetry are (locally) homogeneous.  Indeed, there is a critical
fraction $\half \leq \nu_c \leq \frac34$, so that if $\nu>\nu_c$ then
homogeneity is guaranteed.  It is a natural question to ask whether
the same holds for heterotic backgrounds.  There are examples of
half-BPS heterotic backgrounds which are of cohomogeneity one (see,
e.g., \cite{DKLReview}), hence we know that the critical fraction must
satisfy $\nu_c \geq \half$.

Suppose therefore that we have a background with $\nu>\half$.  In
particular, this means that the space of $D$-parallel spinors must
have dimension $d>8$.  This means that the holonomy group of $D$ must
be contained in the subgroup of $\Spin(1,9)$ which fixes $d>8$
linearly-independent spinors.  The possible stabilizer subgroups of
spinors come in two families
\cite{JMWaves,FKYHeterotic,GLP,GPRSTypeI}, whose Lie algebras are
\begin{equation*}
\fspin(7)\ltimes\RR^8 \supset \fg_2 \supset \fsu(3) \supset \fsp(1)
\end{equation*}
and
\begin{multline*}
\fspin(7)\ltimes\RR^8 \supset \fsu(4) \ltimes \RR^8 \supset \fsp(2)
\ltimes \RR^8\\  \supset (\fsp(1)\oplus\fsp(1))\ltimes \RR^8 \supset
\fsp(1) \ltimes \RR^8 \supset \fu(1)\ltimes \RR^8 \supset \RR^8~,
\end{multline*}
along which the dimension of the subspace of invariant spinors
increases from left to right: being $1$ for $\fspin(7)\ltimes\RR^8$
and $8$ for either $\fsp(1)$ or $\RR^8$, whence if $d>8$ the holonomy
algebra of $D$ must be trivial, so that $D$ must be flat.  The same
group-theoretical argument shows the kernel of the gaugino variation
$Q$ can be at most $8$-dimensional, unless the gauge field is flat.
Therefore if the space of Killing spinors has dimension $>8$, $F=0$.
By the results of \cite{FKYHeterotic} the background must have
constant dilaton and it follows from \cite{JMFPara} that it must be
locally isometric to a Lie group with a bi-invariant lorentzian
metric, whence in particular it is locally homogeneous.  We conclude
that $\nu_c = \half$ for heterotic backgrounds.

It is natural to ask whether also here, as in eleven-dimensional
supergravity, homogeneity is a direct consequence of supersymmetry.
In other words, whether at every point $p \in M$, there is a basis for
$T_pM$ made out of vectors from $[\fg_1,\fg_1]$.  The arguments in
\cite[§6.2]{FMPHom} depend on the symplectic structure of the spinor
representation, which is missing in the case of chiral spinors, but
nevertheless a very similar proof works after some modification.

Let us fix a point $p\in M$ once and for all.  The fibre of the spinor
bundle at $p$ can be identified with the chiral spinor representation
$S_+$, whereas the values at $p$ of the Killing spinors define a
subspace $W \subset S_+$.  The squaring map defines a map $\Phi: W
\otimes W \to T_pM$ and we would like to show that this map is
surjective for $W$ of sufficiently high dimension.  To show that
$\Phi$ surjects onto $T_pM$, it is enough to show that the
perpendicular complement of its image vanishes.  A tangent vector $v
\in T_pM$ is perpendicular to the image of $\Phi$ if and only if for
every $\varepsilon_1,\varepsilon_2 \in W$,
\begin{equation*}
  v^m \overline\varepsilon_1 \gamma_m\varepsilon_2 = 0~.
\end{equation*}
In other words, Clifford multiplication by $v$ maps $W$ to a subspace
of $S_-$ which corresponds to its annihilator $W^0$ under the natural
isomorphism $S_- \cong S_+^*$ explained in Appendix~\ref{app:spinors}.
The existence of $\half$-BPS heterotic backgrounds which are not
homogeneous says that, if there is a critical dimension for $W$ above
which $\Phi$ surjects, this dimension is $>8$.  Therefore without loss
of generality we take $\dim W > 8$.  Since $\dim W + \dim W^0 = 16$,
it follows that $\dim W^0 < 8$ and hence the map $W \to W^0$ defined
by the Clifford action of $v$ has nontrivial kernel for dimensional
reasons.  Since $v^2 = |v|^2$ it follows that $v$ is null, whence, up
to a Lorentz transformation, we can let $v= \be_+$.  The Clifford
relation $\gamma_+ \gamma_- + \gamma_- \gamma_+ = 2$ now says that
$\gamma_+$ has rank $8$.  The rest of the proof now proceeds as in
\cite[§6.2]{FMPHom}.

Indeed, let $U$ be a complementary subspace to $W$, so that $S_+ = W
\oplus U$.  Relative to this split, the symmetric bilinear form
$\beta$ on $S_+$, defined by
\begin{equation*}
  \beta(\varepsilon_1, \varepsilon_2) = \overline\varepsilon_1
  \gamma_- \varepsilon_2~,
\end{equation*}
has the following matrix
\begin{equation*}
  \begin{pmatrix}
    0 & A\\ A^t & B
  \end{pmatrix}~,
\end{equation*}
where $A: U \to W$, $A^t: W \to U$ and $B: U \to U$ are linear maps.
We know that this matrix has rank $8$, since $\gamma_-$ has rank $8$.
As in \cite[§6.2]{FMPHom}, we will now estimate the maximal
possible rank of $\beta$ in terms of the dimension of $W$.

The kernel of $\beta$ consists of $w + u\in W \oplus U$ such that
$Au=0$ and $A^t w + Bu = 0$.  Notice that $\dim U < \dim W$, whence
$\rank A \leq \dim U$.  In the case of maximal rank, the only solution
of $Au=0$ is $u=0$.  In this case the kernel of $\beta$ consists of
$w+0$ with $w\in\ker A^t$.  In other words, the dimensions of the
kernels of $\beta$ and of $A^t$ agree.  Since $A^t$ and $A$ have the
same rank, $A^t$ is onto, whence its kernel has dimension $\dim W -
\dim U$.  Therefore the rank of $\beta$ is \emph{at most} $16 - \dim W
+ \dim U = 2 \dim U$; but we know that the rank of $\beta$ is $8$,
whence $8 \leq 2 \dim U$ or $\dim U \geq 4$.  This means that if
$\dim U < 4$ (equivalently, if $\dim W > 12$) no such $v$ can exist
and the squaring map $\Phi$ is surjective.

Therefore we conclude that if $\nu>\tfrac34$ then the (local)
homogeneity is directly due to the supersymmetry.  It is still an open
question whether this is necessarily the case for homogeneous
backgrounds with $\tfrac12 < \nu \leq \tfrac34$.

\section{Type IIB supergravity backgrounds}
\label{sec:typeIIB}

We now tackle the case of type IIB supergravity backgrounds.  Due to
the proliferation of fields, this is computationally more involved
than the case of heterotic or M-theory backgrounds.  We will find
that, just as in eleven-dimensional supergravity \cite{FMPHom} and in
contrast with the heterotic case treated in §\ref{sec:typeI}, the
critical fraction of supersymmetry above which (local) homogeneity is
guaranteed $\half \leq \nu_c \leq \frac{3}{4}$.

In this section we will introduce the type IIB supergravity
backgrounds and then construct their associated Killing superalgebra.
We then show that the critical fraction for homogeneity obeys $\half
\leq \nu_c \leq \frac34$ and compare our results with the few known
solutions preserving that much supersymmetry.

\subsection{Supersymmetric IIB backgrounds}
\label{sec:IIBtheory}

Type IIB supergravity \cite{SchwarzIIB, SchwarzWestIIB, HoweWestIIB}
is the unique 10-dimensional chiral supergravity theory with $32$
supercharges.  It is also the field theory limit of type IIB
superstring theory.

The bosonic dynamical fields are a ten-dimensional lorentzian metric
$g$, the dilaton $\phi$, the Ramond-Ramond (R-R) gauge potentials
$C^{(0)}$, $C^{(2)}$ and $C^{(4)}$ and the NS-NS 2-form gauge
potential $B$. The axion $C^{(0)}$ and dilaton $\phi$ combine into the
axi-dilaton $\tau=C^{(0)}+i e^{-\phi}$ taking values in the upper
half-plane.  The theory has a global $\SL(2,\RR)$ classical duality
symmetry under which $g$ and $C^{(4)}$ are inert, whereas $B$ and
$C^{(2)}$ transform as a doublet, and $\tau$ transform via fractional
linear transformations on the upper-half plane.

The fermionic fields in the theory are often described as a complex
chiral gravitino $\psi_m$ and a complex antichiral axi-dilatino
$\lambda$.  Under the global $\SL(2,\RR)$ symmetry, all fermionic
fields transform by appropriate phases \cite{TomasSL2R,SenNetwork}
corresponding to a local $\SO(2)$ transformation.

We are interested in bosonic backgrounds; that is, solutions of
the theory where the fermions are set to zero. The equations of
motion for the bosonic fields can be found by varying an $\SL(2,\RR)$
invariant pseudoaction \cite{BBOSduality}.  To write it down, let us
combine the potentials into the following fieldstrengths:
\begin{equation*}
  \begin{aligned}[m]
    H &= dB\\
    G^{(1)}&=dC^{(0)}\\
    G^{(3)}&=dC^{(2)}-C^{(0)} H\\
    G^{(5)}&= dC^{(4)}-\tfrac{1}{2} dB \wedge C^{(2)} +\tfrac{1}{2}
    dC^{(2)}\wedge B~.
  \end{aligned}
\end{equation*}
The RR potential $C^{(4)}$ is constrained so that $G^{(5)}$ is
antiself-dual, in our conventions.  In the string frame, we can write
the pseudoaction as
\begin{multline}
  \label{eq:IIBaction}
  I_{\text{IIB}}=\int d^{10}x \sqrt{-g} \left\{ e^{-2\phi}\left(
R + 4 |d\phi|^2 - \tfrac{1}{2}|H|^2 \right) \right.\\
\left. -\tfrac{1}{2} \left(|G^{(1)}|^2  + |G^{(3)}|^2 +
\tfrac{1}{2} |G^{(5)}|^2\right)\right\}\\ -  \tfrac{1}{2}\int C^{(4)}
\wedge dC^{(2)} \wedge dB~.
\end{multline}
The antiself-duality of $G^{(5)}$ must then be imposed by hand.

When discussing supersymmetry it is convenient to think of the complex
fermions as doublets of real fermions, since supersymmetry does not
act complex linearly.  Therefore the supersymmetry parameters will be
taken to be an $\SO(2)$ doublet of real chiral spinors
$(\varepsilon_1,\varepsilon_2)$.  As a representation of the spin
group, they transform as two copies of the chiral spinor
representation $S_+$.  We will denote this representation by
$\SS_+=S_+ \oplus S_+$.  Such a spinor is a Killing spinor of a
bosonic type IIB background if the supersymmetry variations of the the
fermionic fields vanish.  (Those of the bosonic fields are
automatically zero because in a bosonic background the fermions have
been put to zero.)  The variation of the gravitino gives rise to a
differential equation, whereas the equation coming from varying the
dilatino is algebraic.  For type IIB supergravity, a spinor $\beps =
(\varepsilon_1, \varepsilon_2)$ is Killing if it satisfies
\cite{BdRKRIIB,EmilyIIB}
\begin{equation}
  \label{eq:IIBKS}
  \begin{aligned}[m]
    D_m \beps &:= \nabla_m \beps + \tfrac18 H_{mnp}\gamma^{np}\otimes
    \lambda_3 \beps + \widetilde\Omega\gamma_m \beps = 0\\
    P \beps &:= \left(d\phi +\tfrac12 H \otimes \lambda_3 +
      \Omega\right) \beps = 0~,
  \end{aligned}
\end{equation}
where
\begin{equation}
  \label{eq:Omegabar}  
  \widetilde\Omega = \tfrac18 e^\phi \left(G^{(1)} \otimes\lambda_2 -
    G^{(3)} \otimes \lambda_1 + \half G^{(5)}\otimes \lambda_2
  \right)~,
\end{equation}
and
\begin{equation*}
  \Omega = \gamma^m \widetilde \Omega \gamma_m = e^\phi
  \left( \half G^{(3)}\otimes\lambda_1 - G^{(1)}\otimes\lambda_2
  \right)~,
\end{equation*}
where the $2\times 2$ matrices $\lambda_a$, given by $\lambda_1 =
\sigma_1$, $\lambda_2 = i \sigma_2$ and $\lambda_3 = \sigma_3$ span
$\fsl(2,\RR)$.  Note that in both $D_m$ and $P$ we have isolated the
terms $\widetilde\Omega$ and $\Omega$ coming from the RR fields, which
as we will see below is a useful dintinction to make.

Since the equations defining the Killing spinors are linear, the
Killing spinors generate a real vector space, denoted $\fg_1$, of
dimension $32 \nu$, where $\nu$ is the fraction of the supersymmetry
preserved by the background.

As in the case of heterotic backgrounds, an infinitesimal symmetry of
a IIB background is a Killing vector which in addition to the metric,
also leaves invariant the dilaton, the NS-NS $3$-form $H$ as well
as the RR fields $G^{(i)}$ for $i=1,3,5$.  If we let $\fg_0$ denote
the Lie algebra of infinitesimal symmetries relative to the
restriction of the Lie bracket of vector fields, then in the next
section we will prove the existence of a Lie superalgebra $\fg = \fg_0
\oplus \fg_1$ extending $\fg_0$.

\subsection{The Killing superalgebra}
\label{sec:IIBKSA}

In this section we will do for type IIB backgrounds what was done in
§\ref{sec:IKSA} for heterotic backgrounds.  Namely, we will construct
a Lie superalgebra on the vector superspace $\fg = \fg_0 \oplus \fg_1$
which extends the Lie algebra structure on $\fg_0$.  The additional
brackets on $\fg$ are defined just as in the case of heterotic
backgrounds: the bracket $\fg_0 \otimes \fg_1 \to \fg_1$ is the action
of infinitesimal symmetries on Killing spinors given by the spinorial
Lie derivative, whereas the bracket $\fg_1 \otimes \fg_1 \to \fg_0$ is
the squaring operation on spinors.  As before, there are several
things that need to be proven.  First of all, we will show that the
image of the squaring map is indeed in $\fg_0$; in other words, that
the vector field obtained by squaring a spinor is Killing and in
addition leaves all the other bosonic fields in the background
invariant.  Then we must also show that the Lie derivative of a
Killing spinor along a Killing vector in $\fg_0$ is again a Killing
spinor.  Finally, we must show that the Jacobi identities are
satisfied.  The $(\fg_0,\fg_0,\fg_0)$ identity is precisely the Jacobi
identity for the Lie bracket of vector fields on a manifold.
Similarly, the $(\fg_0,\fg_0,\fg_1)$ and $(\fg_0,\fg_1,\fg_1)$
identities follow from standard properties of the spinorial Lie
derivative, as explained above and in \cite{JMFKilling,FMPHom}.
As usual, it remains to show that the $(\fg_1,\fg_1,\fg_1)$ Jacobi
identity holds, and this will be shown to be the case by calculation.

\subsubsection{Closure}
\label{sec:closure}

Let $\beps=(\varepsilon_1, \varepsilon_2)$ be a IIB Killing spinor.
Let $K$ be the vector field obtained by squaring this spinor; that is,
\begin{equation*}
  K^m := \overline\beps \gamma^m \beps = \overline\varepsilon_1
  \gamma^m \varepsilon_1 + \overline\varepsilon_2 \gamma^m
  \varepsilon_2~.
\end{equation*}
We will show that $K\in\fg_0$; that is, it is a Killing vector which
leaves invariant all the bosonic fields in the background.  These
results can be read off from the results of \cite{EmilyIIB}, but we
record them here for completeness.

Differentiating $K$, using the differential Killing spinor equation
and simplifying using the Clifford algebra, we find
\begin{align*}
  \nabla_m K_n &= \overline{\nabla_m\beps} \gamma_n \beps +
  \overline\beps
  \gamma_n \nabla_m \beps\\
  &= -\half H_{mnp} K^p + \overline\beps (\gamma_m\widetilde\Omega^*
  \gamma_n - \gamma_n \widetilde\Omega \gamma_m)\beps~,
\end{align*}
where $\widetilde\Omega^*$ is the symplectic adjoint of
$\widetilde\Omega$.  It follows from equation~\eqref{eq:adjoints},
that $\widetilde\Omega^* = \widetilde\Omega$, whence
\begin{equation}
  \label{eq:nablaK}
  \nabla_m K_n = -\half H_{mnp} K^p + \overline\beps
  (\gamma_m \widetilde\Omega \gamma_n - \gamma_n \widetilde\Omega
  \gamma_m)\beps = - \nabla_n K_m~,
\end{equation}
showing that $K$ is a Killing vector.

To show that $K$ preserves the dilaton, we use dilatino Killing spinor
equation $P\beps = 0$.  Indeed,
\begin{align*}
  \overline\beps P \beps &= \overline\beps \left(d\phi + \half H \otimes
    \lambda_3 + \Omega \right) \beps\\
  &= \overline \beps d\phi \beps\\
  &= K^m \partial_m\phi~,
\end{align*}
where in the second line we have used the fact that
$H\otimes\lambda_3$ and $\Omega$ are symplectically self-adjoint, and
that for such an endomorphism $A$,
\begin{equation*}
  \overline\beps A \beps = \overline{A\beps}\beps = - \overline\beps A
  \beps \qquad\implies\qquad \overline\beps A \beps = 0~.
\end{equation*}
Therefore we conclude that $K^m\partial_m\phi = 0$ and hence that $K$
preserves the dilaton.

To show that $K$ preserves the axion $C^{(0)}$, we must show that
$K^m\partial_m C^{(0)} = 0$.  To do this we again use the dilatino
Killing spinor equation, but we first multiply it by $\lambda_2$:
\begin{align*}
  \overline\beps \lambda_2 P \beps &= \overline\beps \left(
    d\phi\otimes \lambda_2 - \half H \otimes \lambda_1 + 
    \half e^\phi G^{(3)}\otimes\lambda_3 + e^\phi G^{(1)}\right)
  \beps\\
  &= e^\phi \overline\beps G^{(1)} \beps~,
\end{align*}
where the other terms vanish because of the self-adjointness of the
endomorphisms $d\phi\otimes\lambda_2$, $H\otimes\lambda_1$ and
$G^{(3)}\otimes \lambda_3$.  Therefore $\imath_K G^{(1)} = 0$, which
says that $K^m \partial_m C^{(0)} = 0$ as desired.

To show that $K$ preserves $H$, let us introduce the vector field
$L$ defined by
\begin{equation*}
  L^m := \overline\beps (\gamma^m\otimes\lambda_3) \beps~.
\end{equation*}
Differentiating $L$, using the differential Killing spinor equation
and simplifying using the Clifford algebra, we find
\begin{align*}
  \nabla_m L_n &= \overline{\nabla_m\beps} (\gamma_n\otimes \lambda_3)
  \beps + \overline\beps (\gamma_n\otimes\lambda_3) \nabla_m \beps\\
  &= -\half H_{mnp} K^p + \overline\beps
  \left(\gamma_m\widetilde\Omega (\gamma_n\otimes\lambda_3) -
    (\gamma_n\otimes\lambda_3) \widetilde\Omega \gamma_m\right)\beps~,
\end{align*}
where we have used that $\widetilde\Omega^* = \widetilde\Omega$.
Noticing that $\widetilde\Omega \lambda_3 = - \lambda_3
\widetilde\Omega$, we can rewrite this as
\begin{equation*}
  \nabla_m L_n = -\half H_{mnp} K^p + \overline\beps
  \left(\gamma_m\widetilde\Omega (\gamma_n\otimes\lambda_3) +
    \gamma_n\widetilde\Omega (\gamma_m\otimes\lambda_3\right) \beps~.
\end{equation*}
Antisymmetrizing, we find
\begin{equation*}
  \nabla_m L_n  - \nabla_n L_m  = - H_{mnp} K^p~,
\end{equation*}
which in the language of differential forms becomes
\begin{equation}
  \label{eq:K11K22}
  \imath_K H = -dL~.
\end{equation}
Since $dH = 0$, this shows that $\eL_K H = 0$.

Similar but more involved calculations, which have thus been relegated
to Appendix~\ref{app:KG3G5}, show that $K$ also preserves $G^{(3)}$
and $G^{(5)}$.  Therefore $K \in \fg_0$.

The proof that the Lie derivative of a Killing spinor along an
infinitesimal symmetry $K \in \fg_0$ is again a Killing spinor is
basically the same as the one for the heterotic backgrounds: it uses
equation~\eqref{eq:Lnabla} and the fact that $K$ preserves the other
fields in the background to deduce that
\begin{equation*}
  \eL_K D_X \beps = D_X \eL_K \beps + D_{[K,X]} \beps,
\end{equation*}
whence if $\beps$ satisfies the differential Killing spinor equation
so does $\eL_K \beps$.  Similarly, since $P$ is invariant under $K$,
if $\beps$ obeys the algebraic Killing spinor equation, $P\beps = 0$,
then so does $\eL_K\beps$.  In summary, $\eL_K\beps$ is again a
Killing spinor.

\subsubsection{The Jacobi identities}
\label{sec:IIBjac}

We now turn our attention to proving the Jacobi identities.  As usual,
all Jacobi identities except for the $(\fg_1,\fg_1,\fg_1)$ identity
follows from general geometric principles.  The proof of the remaining
Jacobi identity requires a calculation.  Polarising this identity, we
see that it is enough to show that for all $\beps \in \fg_1$,
\begin{equation}
  \label{eq:ooojac}
  \eL_K \beps=0 ~,
\end{equation}
where $K^m=\overline\beps\gamma^m\beps$.  We prove this now.

The spinorial Lie derivative is given by
\begin{equation*}
  \eL_K \beps = K^m \nabla_m \beps + \tfrac14 \nabla_mK_n
  \gamma^{mn}\beps~.
\end{equation*}
We now use the differential Killing spinor equation in order to
express $\nabla_m\beps$ in terms of algebraic operations on the
spinor:
\begin{equation*}
  \nabla_m \beps = - \tfrac18 H_{mnp}\gamma^{np}\otimes\lambda_3 \beps
  - \widetilde\Omega \gamma_m \beps~,
\end{equation*}
and equation \eqref{eq:nablaK} for $\nabla_mK_n$.  Using the Clifford
algebra and simplifying, we arrive at
\begin{multline*}
  \eL_K \beps = -\tfrac18 K^m H_{mnp}\gamma^{np}\otimes \lambda_3\beps
  - \tfrac18 H_{mnp}\overline\beps\gamma^p \otimes \lambda_3\beps
  \gamma^{mn}\beps\\
  - \widetilde\Omega K^m \gamma_m \beps + \half \overline\beps
  \gamma_m \widetilde\Omega \gamma_n \beps \gamma^{mn}\beps~.
\end{multline*}
Breaking $\beps = (\varepsilon_1, \varepsilon_2)$ into components and
expanding $K^m = \overline\beps \gamma^m\beps$, the $H$-dependent
terms become
\begin{equation}
  \label{eq:Hdep}
  \begin{pmatrix}
    -\tfrac14 H_{mnp} \overline\varepsilon_1 \gamma^m \varepsilon_1
    \gamma^{np}\varepsilon_1\\[5pt]
    \phantom{+}\tfrac14 H_{mnp} \overline\varepsilon_2 \gamma^m \varepsilon_2
    \gamma^{np}\varepsilon_2
  \end{pmatrix}~.
\end{equation}

Expanding the RR-terms and using that $\overline\varepsilon_i \gamma^m
\varepsilon_i \gamma_m \varepsilon_i = 0$ for $i=1,2$, we obtain
\begin{equation*}
  \tfrac18 e^\phi
  \begin{pmatrix}
    -\overline\varepsilon_1 \gamma^m\varepsilon_1 \GG^- \gamma_m \varepsilon_2 +
    \overline\varepsilon_2 \gamma_n \GG^+ \gamma_m
    \varepsilon_1\gamma^{mn}\varepsilon_1\\[5pt]
    \overline\varepsilon_2 \gamma^m\varepsilon_2 \GG^+ \gamma_m \varepsilon_1 +
    \overline\varepsilon_2 \gamma_n \GG^+ \gamma_m
    \varepsilon_1\gamma^{mn}\varepsilon_2    
  \end{pmatrix}~,
\end{equation*}
where $\GG^\pm = G^{(1)} \pm G^{(3)} + \half G^{(5)}$.  This can be
simplified further using the Fierz identity
\begin{multline*}
  -\overline\alpha_1 \gamma_m G^{(k)} \gamma_n \alpha_2
  \gamma^{mn}\alpha_1 + \overline\alpha_1 \gamma_m \alpha_1 G^{(k)}
  \gamma^m \alpha_2 \\
  = (10-2k) \left( \overline\alpha_1 G^{(k)}\alpha_2
    \alpha_1 - \half \overline\alpha_1\gamma^m\alpha_1 \gamma_m
    G^{(k)} \alpha_2\right)~,
\end{multline*}
for any two positive chirality spinors $\alpha_i$, which is proved in
Appendix~\ref{app:fierz} as equation \eqref{eq:fierzG}.  Indeed, using
this identity we first of all see that the terms involving the RR
$5$-form vanish identically, whereas the rest becomes after some
simplification
\begin{equation}
  \label{eq:RRprelim}
  e^\phi
  \begin{pmatrix}
    \overline\varepsilon_1 \left( \half G^{(3)} - G^{(1)}\right)
    \varepsilon_2\varepsilon_1 - \half \overline\varepsilon_1\gamma^m
    \varepsilon_1 \gamma_m \left( \half G^{(3)} - G^{(1)}\right)
    \varepsilon_2 \\[5pt]
    \overline\varepsilon_2 \left( \half G^{(3)} + G^{(1)}\right)
    \varepsilon_1\varepsilon_2 - \half \overline\varepsilon_2\gamma^m
    \varepsilon_2 \gamma_m \left( \half G^{(3)} + G^{(1)}\right)
    \varepsilon_1    
  \end{pmatrix}~.
\end{equation}
We simplify this further using the algebraic Killing spinor equation,
which in components reads
\begin{align*}
  e^\phi\left( \half G^{(3)} - G^{(1)}\right) \varepsilon_2 &=
  -\left(\half H + d\phi\right) \cdot \varepsilon_1\\
  e^\phi\left( \half G^{(3)} + G^{(1)}\right) \varepsilon_1 &=
  \left(\half H - d\phi\right) \cdot \varepsilon_2~.
\end{align*}
Performing the substitution and after some simplification, the
expression in \eqref{eq:RRprelim} becomes
\begin{equation*}
  \begin{pmatrix}
    -\overline\varepsilon_1 d\phi\varepsilon_1 \varepsilon_1 + \half
    \overline\varepsilon_1\gamma^m \varepsilon_1 \gamma_m
    d\phi\varepsilon_1 + \tfrac14 \overline\varepsilon_1 \gamma^m
    \varepsilon_1\gamma_m H \varepsilon_1 \\[5pt]
    -\overline\varepsilon_2 d\phi\varepsilon_2 \varepsilon_2 + \half
    \overline\varepsilon_2\gamma^m \varepsilon_2 \gamma_m
    d\phi\varepsilon_2 - \tfrac14 \overline\varepsilon_2 \gamma^m
    \varepsilon_2\gamma_m H \varepsilon_2 
  \end{pmatrix}~.
\end{equation*}
It is easy to see that the dilaton-dependent terms vanish.  Indeed,
for any positive-chirality spinor $\alpha$,
\begin{equation*}
  -\overline\alpha d\phi \alpha \alpha + \half \overline\alpha
  \gamma^m \alpha \gamma_md\phi \alpha = 0~,
\end{equation*}
using the Clifford algebra in the second term, together with the
identity $\overline\alpha\gamma^m\alpha\gamma_m\alpha = 0$.
Adding the contribution in equation \eqref{eq:Hdep} to the remaining
terms, we find
\begin{equation*}
  4 \eL_K \beps =
  \begin{pmatrix}
    \overline\varepsilon_1 \gamma^m \varepsilon_1 \gamma_m H
    \varepsilon_1 -  \overline\varepsilon_1 \gamma^m \varepsilon_1
    H_{mnp}\gamma^{np} \varepsilon_1\\[5pt]
    -\overline\varepsilon_2 \gamma^m \varepsilon_2 \gamma_m H
    \varepsilon_2 -  \overline\varepsilon_2 \gamma^m \varepsilon_2
    H_{mnp}\gamma^{np} \varepsilon_2   
  \end{pmatrix}~,
\end{equation*}
which is again seen to cancel after using the Clifford algebra and the
identity $\overline\varepsilon_i \gamma^m \varepsilon_i \gamma_m
\varepsilon_i = 0$ for $i=1,2$.

\subsection{Homogeneity of IIB backgrounds}
\label{sec:IIBHom}

In this section we consider the fraction $\nu_c$ of supersymmetry
above which local homogeneity is guaranteed.  The classification
\cite{FOPMax} of maximally supersymmetric solutions and the
nonexistence of preonic solutions \cite{NoIIBPreons} puts $\nu_c\leq
\tfrac{15}{16}$.  In this section we will show that just as in
eleven-dimensional supergravity $\nu_c\leq \tfrac34$.  Indeed, we will
show that if $\nu>\tfrac34$ then the background is locally
homogeneous.  In other words, we will show that at every point $p \in
M$, there is a basis for $T_pM$ made out of vectors from
$[\fg_1,\fg_1]$.  The proof is virtually identical to that of the
heterotic backgrounds.

Let us fix a point $p\in M$ once and for all.  The fibre of the spinor
bundle at $p$ can be identified with the chiral spinor representation
$\SS_+$, whereas the values at $p$ of the Killing spinors define a
subspace $W \subset \SS_+$.  The squaring map defines a map $\Phi: W
\otimes W \to T_pM$ and we would like to show that this map is
surjective for $W$ of sufficiently high dimension.  To show that
$\Phi$ surjects onto $T_pM$, it is enough to show that the
perpendicular complement of its image vanishes.  A tangent vector $v
\in T_pM$ is perpendicular to the image of $\Phi$ if and only if for
every $\varepsilon_1,\varepsilon_2 \in W$,
\begin{equation*}
  v^m \overline\varepsilon_1 \gamma_m\varepsilon_2 = 0~.
\end{equation*}
In other words, Clifford multiplication by $v$ maps $W$ to a subspace
of $\SS_-$ which corresponds to its annihilator $W^0$ under the
natural isomorphism $\SS_- \cong \SS_+^*$ explained in
Appendix~\ref{app:spinors}.  The existence of the cohomogeneity-one
$\half$-BPS D3-brane background shows that, if there is a critical
dimension for $W$ above which $\Phi$ surjects, this dimension is $>16$.
Therefore without loss of generality we take $\dim W > 16$.  Since
$\dim W + \dim W^0 = 32$, it follows that $\dim W^0 < 16$ and hence the
map $W \to W^0$ defined by the Clifford action of $v$ has nontrivial
kernel for dimensional reasons.  Since $v^2 = |v|^2$ it follows that
$v$ is null, whence, up to a Lorentz transformation, we can let $v=
\be_+$.  The Clifford relation $\gamma_+ \gamma_- + \gamma_- \gamma_+
= 2$ now says that $\gamma_+$ has rank $16$.  The rest of the proof now
proceeds as in \cite[§6.2]{FMPHom} or as in §~\ref{sec:typeIHom}.

Indeed, let $U$ be a complementary subspace to $W$, so that $\SS_+ = W
\oplus U$.  Relative to this split, the symmetric bilinear form
$\beta$ on $\SS_+$, defined by
\begin{equation*}
  \beta(\varepsilon_1, \varepsilon_2) = \overline\varepsilon_1
  \gamma_- \varepsilon_2~,
\end{equation*}
has the following matrix
\begin{equation*}
  \begin{pmatrix}
    0 & A\\ A^t & B
  \end{pmatrix}~,
\end{equation*}
where $A: U \to W$, $A^t: W \to U$ and $B: U \to U$ are linear maps.
We know that this matrix has rank $16$, since $\gamma_-$ has rank $16$.
As in \cite[§6.2]{FMPHom}, we will now estimate the maximal
possible rank of $\beta$ in terms of the dimension of $W$.

The kernel of $\beta$ consists of $w + u\in W \oplus U$ such that
$Au=0$ and $A^t w + Bu = 0$.  Notice that $\dim U < \dim W$, whence
$\rank A \leq \dim U$.  In the case of maximal rank, the only solution
of $Au=0$ is $u=0$.  In this case the kernel of $\beta$ consists of
$w+0$ with $w\in\ker A^t$.  In other words, the dimensions of the
kernels of $\beta$ and of $A^t$ agree.  Since $A^t$ and $A$ have the
same rank, $A^t$ is onto, whence its kernel has dimension $\dim W -
\dim U$.  Therefore the rank of $\beta$ is \emph{at most} $32 - \dim W
+ \dim U = 2 \dim U$; but we know that the rank of $\beta$ is $16$,
whence $16 \leq 2 \dim U$ or $\dim U \geq 8$.  This means that if
$\dim U < 8$ (equivalently, if $\dim W > 24$) no such $v$ can exist
and the squaring map $\Phi$ is surjective.

As in the case of eleven-dimensional supergravity, all known IIB
backgrounds with $\nu>\half$ are homogeneous, whence were it not for
the above result, an empirically reasonable conjecture might be that
$\nu_c=\half$.

Indeed, type IIB backgrounds with 24 supersymmetries have been given
in \cite{Michelson, CLPpp, CLPppM, ChrisJerome, BenaRoiban}.  In
\cite{Michelson} a 24 supersymmetric background was constructed by
discrete quotients of the maximally supersymmetric pp-wave
\cite{NewIIB}.  The resulting background inherits the
homogeneity.  In \cite{CLPpp} (see also \cite{CLPppM, ChrisJerome}) a
homogeneous pp-wave with 24 supersymmetries is given that can be
obtained as a Penrose limit of a D3/D3 intersection.  Only the
$5$-form is turned on, but it has a different structure to the
maximally supersymmetric one.  In \cite{BenaRoiban} there are two
families of homogeneous pp-waves with 24 supersymmetries.

The above examples are all homogeneous plane waves.  The homogeneous 
plane-wave metrics are classified in \cite{BOLhpw}.   It is proven in
\cite[§3.1]{BMO} that eleven-dimensional plane waves admitting more than
$16$ supersymmetries are automatically homogeneous.  Their argument is
in fact more general and it is not hard to modify it to show that IIB
plane waves with more than $16$ supersymmetries are homogeneous too.

\section{Conclusion}
\label{sec:conclusion}

In this paper we have constructed the Killing superalgebra for
supersymmetric heterotic and type IIB backgrounds and have shown that
it is indeed a Lie superalgebra, extending the similar result in the
eleven-dimensional case.  We have also shown that any heterotic
background with more than $8$ supersymmetries is locally homogeneous.
Our proof uses the knowledge of the possible holonomy groups of the
metric connection with torsion appearing in the differential Killing
spinor equation.  This is a result which is missing in the case of IIB
supergravity, whence in this case we have to argue, as was done in
\cite{FMPHom} for eleven-dimensional supergravity, that the Killing
vectors constructed from the Killing spinors act locally transitively
provided that the background has more than $24$ supersymmetries.  The
same method of proof would also show that at least $13$
supersymmetries are needed in the case of heterotic supergravity.
Since as we now know this result is sub-optimal, it suggests that our
proof for type IIB is not yielding a sharp result and that the
critical fraction for (local) homogeneity is actually lower than
$\frac34$.  The same could be said about eleven-dimensional
supergravity, of course.  Were it not for these results, the available
evidence would suggest that the critical fraction is actually $\nu_c =
\half$.

What about other ten-dimensional supergravities?  Type I is formally
identical to the heterotic case, but ignoring the gauge fields.  In
particular, since the differential Killing spinor equation is the
same, it is easy to see that the same results apply: local homogeneity
is guaranteed for backgrounds admitting more than $8$ supersymmetries.

The type IIA results follow from those in \cite{FMPHom} and show that
$24+$ implies local homogeneity.  Indeed, a $24+$ IIA background
oxidises to a $24+$ background of eleven-dimensional supergravity,
which we know is locally homogeneous.  The eleven-dimensional geometry
is the total space of a locally trivial fibre bundle over the IIA
geometry.  The Killing spinors of the eleven-dimensional supergravity
background obtained via oxidation are constant on the fibres, whence
so are the Killing vectors obtained by squaring them.  We know that
these act locally transitively and therefore their push-downs to the
base also act locally transitively.  This shows that the IIA
background is locally homogeneous.

\section*{Acknowledgments}

We take pleasure in thanking Philipp Lohrmann for the motivation to
treat the heterotic case and George Papadopoulos for useful email
correspondence and for pointing out a sign error in an early draft of
the heterotic case.  EHJ is supported by an EPSRC Postdoctoral
Fellowship.

\appendix

\section{Spinors in $9+1$ dimensions}
\label{app:spinors}

In this appendix we set our spinorial conventions and record a number
of results which are used repeatedly throughout the paper.

We work in signature $(9,1)$ with a mostly plus metric.  The Clifford
product is such that $\bx \cdot \bx = + |\bx|^2$, whence we are
working with $\Cl(9,1)$.  We let $\gamma_m$ denote the corresponding
gamma matrices, obeying
\begin{equation*}
  \gamma_m \gamma_n + \gamma_n \gamma_m = + 2 \eta_{mn}~, 
\end{equation*}
with $\eta_{mn}$ the mostly plus lorentzian inner product.
As a real associative algebra, $\Cl(9,1) \cong \Mat(32,\RR)$, whence
there is a unique irreducible Clifford module $S$, which is real and
of dimension $32$.  Under $\Spin(9,1)$, it decomposes as a direct sum
$S = S_+ \oplus S_-$ of the ($\pm1$)-eigenspaces of the volume form
$\omega$, which obeys $\omega^2 = 1$.

The natural invariant spinor inner product on $S$ is symplectic,
relative to which $S_\pm$ are lagrangian subspaces.  Therefore as
representations of $\Spin(9,1)$, $S_+$ and $S_-$ are naturally dual.
The spin group acts on them with determinant one, but preserves no
nondegenerate bilinear form \cite{Harvey}.  We will write this inner
product as
\begin{equation*}
  \left(\varepsilon_1,\varepsilon_2 \right) := \varepsilon_1^t C
  \varepsilon_2 = \overline\varepsilon_1 \varepsilon_2~,
\end{equation*}
for $\varepsilon_i \in S$, where $C$ is the charge conjugation matrix,
which is antisymmetric.  The action of the Clifford algebra is such
that 
\begin{equation*}
  \left( \gamma_m \varepsilon_1 , \varepsilon_2 \right) = - 
  \left(\varepsilon_1 ,  \gamma_m \varepsilon_2 \right)~,
\end{equation*}
or, equivalently, $C \gamma_m  = - \gamma_m^t C$.  It follows from
this property that the symplectic form is invariant under the spin
group.

As representations of the spin group, the tensor square $S^{\otimes2}$
of the spinor representation is isomorphic to the exterior algebra of
the vector representation.  In particular, for the chiral spinor
representation $S_+$, we find
\begin{equation*}
  S_+ \otimes S_+ \cong \Lambda^1 \oplus \Lambda^3 \oplus
  \Lambda^5_+,
\end{equation*}
of which the symmetric part is $\Symm^2 S_+ \cong \Lambda^1 \oplus
\Lambda^5_+$ and the skewsymmetric part $\Lambda^2S_+ \cong
\Lambda^3$.  Globalising on a manifold, we obtain a ``squaring'' map
from spinor fields to differential forms.  Hence squaring a chiral
spinor field $\varepsilon$ we obtain a $1$-form $K$ and a self-dual
$5$-form $\Xi$, defined explicitly by 
\begin{equation*}
  \begin{aligned}[m]
    K_m &= \overline\varepsilon \gamma_m \varepsilon\\
    \Xi_{m_1\dots m_5} &= \overline\varepsilon \gamma_{m_1\dots m_5}
    \varepsilon~.
  \end{aligned}
\end{equation*}
Under the Clifford product, these forms annihilate the spinor
$\varepsilon$.  This can be shown directly via Fierzing or
equivalently via the following representation-theoretic argument.  The
stabilizer $G_\varepsilon < \Spin(9,1)$ of a nonzero chiral spinor
$\varepsilon$ is isomorphic to $\Spin(7) \ltimes \RR^8$, where in
terms of Lorentz transformations, the $\RR^8$ subgroup consists of
null rotations around a lightlike direction.  This subgroup (or a
slight variant in eleven dimensions) has been explained in detail in
\cite{JMWaves}.  Because the forms $K$ and $\Xi$ are constructed out
of $\varepsilon$, the stabilizer $G_\varepsilon$ also leaves them
invariant.  The invariant forms under $G_\varepsilon$ are easily
determined from the results in \cite[Appendix~A]{JMWaves}, by ignoring
the ninth spatial direction $e_9$.  We see that there is an invariant
$1$-form which is null and is precisely the one along which the null
rotations are defined.  In our notation, this is the one-form $K$.
Under a typical such null rotation in the Lie algebra of
$G_\varepsilon$, $K^m \gamma_i\gamma_m \varepsilon = 0$, where $i$ is
a spacelike direction orthogonal to $K$.  Since $\gamma_i$ is
invertible, it follows that $K^m \gamma_m \varepsilon = 0$.
Similarly, the self-dual $5$-form takes the form $\Xi = K \wedge
\Phi$, with $\Phi$ a Cayley $4$-form, and thus the Clifford product
$\Xi \cdot \varepsilon = 0$ because $K \cdot \varepsilon = 0$.

In IIB supergravity, the relevant spinor representation is isomorphic
to two copies of $S_+$, denoted $\SS_+ := S_+ \oplus S_+$ in this
paper.  The invariant symplectic structure on $S$ extends trivially to
$\SS = S \oplus S$, relative to which $\SS_+$ and $\SS_-:=S_- \oplus
S_-$ are complementary lagrangian subspaces and hence naturally dual.
There is natural action of $\Spin(9,1) \times \GL(2,\RR)$ on $\SS_+$,
where $\Spin(9,1)$ acts separately on each summand and $\GL(2,\RR)$
acts in the obvious way on pairs of spinors, but only $\Spin(9,1)
\times \O(2)$ preserves the symplectic structure.  The classical
duality group $\SL(2,\RR)$ acts via local $\SO(2)$ transformations.

Many of the calculations performed in this paper require working out
the symplectic adjoint of operators acting on $\SS$.  These operators
are a mixture of the Clifford and $\fgl(2,\RR)$ actions on $\SS$.
Typically they will be linear combinations of the monomials of the
form $\varphi \otimes \lambda$, where $\varphi$ is a $p$-form and
$\lambda$ a $2\times 2$ matrix.  The symplectic adjoint $\varphi^*$ of
a $p$-form obeys
\begin{equation*}
  \varphi^* = (-1)^{p(p+1)/2} \varphi~,
\end{equation*}
whereas the adjoint of a $2\times 2$ matrix $\lambda$ is its
transpose.  This allows us to compute easily the symplectic adjoints
of many of the terms which appear in the Killing spinor equations.
Let $\lambda_1 = \sigma_1$, $\lambda_2 = i\sigma_2$ and $\lambda_3 =
\sigma_3$, and let $G^{(p)}$, $p=1,3,5$, be the RR $p$-form
fieldstrengths.  Then for the terms appearing in the Killing spinor
equations we have
\begin{equation}
  \label{eq:adjoints}
  \begin{aligned}[m]
    \left(G^{(1)}\otimes\lambda_2\right)^* &= G^{(1)}\otimes\lambda_2\\
    \left(G^{(3)}\otimes\lambda_1\right)^* &= G^{(3)}\otimes\lambda_1\\
    \left(G^{(5)}\otimes\lambda_2\right)^* &= G^{(5)}\otimes\lambda_2\\ 
    \left(H\otimes\lambda_3\right)^* &= H\otimes\lambda_3~,
  \end{aligned}
\end{equation}
whence $\widetilde\Omega^* = \widetilde\Omega$ and $\Omega^* =
\Omega$.

\section{$K$ preserves $G^{(3)}$ and $G^{(5)}$}
\label{app:KG3G5}

In this appendix we show that $\eL_KG^{(3)} =0$ and $\eL_K G^{(5)}
=0$, where $K^m = \overline\beps\gamma^m\beps$ is the Killing vector
obtained by squaring a Killing spinor $\beps$.  In the main text we
have already shown that $K$ preserves $H$ and the axi-dilaton.

To prove that $H$ preserves $G^{(3)}$ and $G^{(5)}$, we will use
identities derived in \cite{EmilyIIB} for the differential forms
constructed out of a Killing spinor.  Let us define the following
differential forms:
\begin{align*}
  (k_{ij})_m &= \bar\varepsilon_i \gamma_m \varepsilon_j\\
  (\xi^{(3)}_{ij})_{mnr} &=
  \bar{\varepsilon}_i\gamma_{mnr}\varepsilon_j\\
  (\xi^{(5)}_{ij})_{mnrpq} &=
  \bar{\varepsilon}_i\gamma_{mnrpq}\varepsilon_j~,
\end{align*}
where $\beps=(\varepsilon_1,\varepsilon_2)$.  In this language, the
Killing vector is $K=k_{11} + k_{22}$ and the vector $L$ introduced in
§\ref{sec:closure} is given by $L = k_{11} - k_{22}$.  We will use the
notation $\Xi:=\xi^{(5)}_{11} + \xi^{(5)}_{22}$ and $\Theta:=
\xi^{(5)}_{11} - \xi^{(5)}_{22}$.  As explained in
Appendix~\ref{app:spinors}, the $1$-forms and $5$-forms are symmetric
under $i\leftrightarrow j$, whereas the 3-form is antisymmetric.

The following identity is a direct  consequence of the differential
Killing spinor equation \cite{EmilyIIB}:
\begin{equation}
  \label{eq:K12}
  (dk_{12})_{mn} = \tfrac12 H^{rs}{}_{[m} (\xi^{(3)}_{12})_{n]rs} +
  \tfrac14 e^{\phi}\left( \imath_K G^{(3)} + \imath_{G^{(3)}}
    \Xi\right)_{mn}~.
\end{equation}
We can manipulate this identity using the algebraic Killing spinor
equation.  In particular, the identity $\overline\beps
\gamma_{mn}\otimes \lambda_1 P \beps= 0$ yields
\begin{equation*}
  2 k_{12} \wedge d\phi + e^{\phi} L \wedge G^{(1)}
  - (\xi^{(3)}_{12})_{rs[m} H_{n]}{}^{rs} + \half e^{\phi}
  \imath_{G^{(3)}} \Xi - \half e^{\phi} \imath_K G^{(3)} = 0~,
\end{equation*}
which can be inserted into \eqref{eq:K12} to yield
\begin{equation*}
  d k_{12} = - k_{12}\wedge d\phi - \half e^{\phi} L \wedge G^{(1)} +
  \half e^{\phi} \imath_K G^{(3)}~,
\end{equation*}
or, equivalently,
\begin{equation*}
  2 d(e^{-\phi}k_{12}) = -L \wedge G^{(1)} + \imath_K G^{(3)}~.
\end{equation*}
Differentiating again, we see that
\begin{equation*}
  d(\imath_K G^{(3)}) = dL \wedge G^{(1)}~.
\end{equation*}
On the other hand, using the definition of $G^{(3)}$ and the fact that
$\imath_K G^{(1)} =0$, we also have
\begin{equation*}
  \imath_K d G^{(3)} = \imath_K(H\wedge G^{(1)}) = \imath_K H \wedge
  G^{(1)} = - dL\wedge G^{(1)}~.
\end{equation*}
Combining the two results, we see that
\begin{equation*}
  \eL_K G^{(3)} =  \imath_K d G^{(3)} + d\imath_K G^{(3)} = 0~.
\end{equation*}

For $G^{(5)}$ we follow a similar procedure.  The differential
equation for the 3-form $\xi^{(3)}_{12}$ arising from the differential
Killing spinor equation is
\begin{multline*}
    (d \xi^{(3)}_{12})_{mnpq} =
    \half e^{\phi} \left( \imath_{G^{(1)}} \Xi - \imath_K
    G^{(5)} + \half L \wedge G^{(3)} - G^{(3)}_{rs[m}\Theta_{npq]}{}^{rs}\right)\\
  + \tfrac{3}{2} k_{12} \wedge H + H^{rs}{}_{[m} (\xi^{(5)}_{12})_{npq]rs}~.
\end{multline*}
We can simplify this using the algebraic Killing spinor
equation. Indeed, the equation $\overline\beps \gamma_{mnpq}\otimes
\lambda_2 P \varepsilon =0$ can be used to rewrite the
above equation as
\begin{equation*}
  d \xi^{(3)}_{12} = k_{12}\wedge H + d\phi \wedge \xi^{(3)}_{12} +
  \half e^{\phi} L \wedge G^{(3)} - \half e^{\phi}\imath_{K}
  G^{(5)}~.
\end{equation*}
Differentiating and resubstituting the expressions for
$d\xi^{(3)}_{12}$ and $dk_{12}$, gives
\begin{equation*}
  dL \wedge G^{(3)} - d(\imath_K G^{(5)}) + \imath_K G^{(3)} \wedge H
  = 0~.
\end{equation*}
Now, we also have
\begin{equation*}
  \imath_K d G^{(5)} = \imath_K( H\wedge G^{(3)})  = -dL \wedge
  G^{(3)} - H\wedge \imath_K G^{(3)}~,
\end{equation*}
where we have used \eqref{eq:K11K22} to replace $\imath_K H$.
Combining the above two equations gives
\begin{equation*}
  \eL_K G^{(5)} = \imath_K d G^{(5)} + d\imath_K G^{(5)} = 0~.
\end{equation*}

\section{A Fierz identity}
\label{app:fierz}

Let $S=S_+ \oplus S_-$ denote the unique irreducible spinor module of
$\Cl(9,1)$.  As recalled in Appendix~\ref{app:spinors}, it is real,
symplectic and of dimension $32$.  Any two nonzero spinors
$\alpha,\beta\in S$ define a rank-$1$ endomorphism $\alpha\otimes
\overline\beta: S \to S$ by
\begin{equation*}
  (\alpha\otimes \overline\beta)(\varepsilon) =
  (\overline\beta\varepsilon)\alpha~.
\end{equation*}
Since $\Cl(9,1) \cong \End(S)$, there is a unique element of $\Cl(9,1)$
corresponding to this endomorphism.  Taking traces, it is not
difficult to derive the following expression for this element:
\begin{equation}
  \label{eq:fierz}
  \alpha\otimes\overline\beta = \tfrac{1}{32} \sum_{k=0}^{10}
  \tfrac{1}{k!} (-1)^{k(k-1)/2} \overline\beta\gamma^{a_1\cdots
    a_k}\alpha \gamma_{a_1\cdots a_k}~,
\end{equation}
where we have an implicit summation over the $a_i$ indices and where
$\gamma^{a_1\cdots a_k}$ is equal to the identity for $k=0$.  We will
need the special case of this identity when $\alpha,\beta$ have
definite chirality.

We record here two useful Clifford identities.  Firstly, the Clifford
product of a $p$-form $\theta$ and the volume form $\omega$ defining
the notion of chirality is given by
\begin{equation}
  \label{eq:dvol}
  \theta \cdot \omega = (-1)^{p(p-1)/2} \star\theta~,
\end{equation}
where $\star\theta$ is the Hodge dual defined by
\begin{equation*}
  \theta \wedge \star\theta = |\theta|^2 \omega~.
\end{equation*}
Similarly, for $\theta$ a $p$-form,
\begin{equation}
  \label{eq:trace}
  \gamma^m \theta \gamma_m = (-1)^p (10-2p) \theta~.
\end{equation}

Using identity \eqref{eq:dvol}, we may derive a master Fierz identity
for three spinors $\alpha_i$, $i=1,2,3$, with definite chiralities
$\chi_1, \chi_2, \chi_3$, with $\chi_2=-\chi_1$:
\begin{equation}
  \label{eq:fierzchiral}
  (\overline\alpha_1\alpha_2)\alpha_3 =
  \tfrac{1}{32} \sum_{n=0}^5 a_{n(\chi_1 \chi_3)}
  \tfrac{1}{n!}\left(\overline\alpha_1 \gamma_{c_1\cdots c_n} \alpha_3
  \right) \gamma^{c_1 \cdots c_n} \alpha_2~,
\end{equation}
where the coefficients $a_{n(\chi_1 \chi_3)}$ are determined as follows
\begin{equation*}
  \renewcommand{\arraystretch}{1.2}
  \begin{tabular}{|*{3}{>{$}c<{$}|}}
    \hline
    n & a_{n(\pm \pm)} & a_{n(\pm\mp)} \\
    \hline
    0 & 0 & 2\\
    1 & 2 & 0\\
    2 & 0 & -2\\
    3 &-2 & 0\\
    4 & 0 & 2\\
    5 & 1 & 0\\\hline
  \end{tabular}
\end{equation*}
Finally, we use this to derive an identity used in §\ref{sec:IIBjac}
to prove the Jacobi identity for the Killing superalgebra.

Let $\alpha_{1,2}$ be two positive-chirality spinors and let
$G^{(k)}$ be an $k$-form, where $k$ is odd.  Then the following
identity holds
\begin{multline}
  \label{eq:fierzG}
  \overline\varepsilon_1 \gamma_m G^{(k)} \gamma_n \varepsilon_2
  \gamma^{nm} \varepsilon_1 + \overline\varepsilon_1 \gamma_m
  \varepsilon_1 G^{(k)} \gamma^{m} \varepsilon_2\\
  = - (-1)^k(10-2k) \left( \overline\varepsilon_1 G^{(k)}
    \varepsilon_2 \varepsilon_1 + \half \overline\varepsilon_1
    \gamma_m \varepsilon_1 \gamma^m G^{(k)}\varepsilon_2 \right)~.
\end{multline}

To prove this identity, we begin by fierzing
$\overline\varepsilon_1\gamma_m G^{(k)} \gamma_n\varepsilon_2 \gamma^n
\gamma^m \varepsilon_1$.  We do this by using the master Fierz
identity \eqref{eq:fierzchiral} with
$\overline\alpha_1=\overline\varepsilon_1 \gamma_m$,
$\alpha_2=G^{(k)}\gamma_n\varepsilon_2$ and
$\alpha_3=\gamma^n\gamma^m\varepsilon_1$.   Notice that in this case,
$\chi_1=\chi_3=+1$, whence
\begin{multline*}
  \overline\varepsilon_1 \gamma_m G^{(k)} \gamma_n \varepsilon_2
  \gamma^n \gamma^m \varepsilon_1 = \tfrac1{16}
  \overline\varepsilon_1 \gamma_m \gamma^n \gamma^m \varepsilon_1
  G^{(k)} \gamma_n\varepsilon_2\\
  - \tfrac1{32}\overline\varepsilon_1 \gamma_m \gamma_{pq}\gamma^n
  \gamma^m \varepsilon_1 \gamma^{pq}G^{(k)}\gamma_n\varepsilon_2 +
  \tfrac{1}{384} \overline\varepsilon_1 \gamma_m \gamma_{pqrs}\gamma^n
  \gamma^m \varepsilon_1 \gamma^{pqrs} G^{(k)}\gamma_n\varepsilon_2~.
\end{multline*}
The third term gives zero using the trace identity \eqref{eq:trace}
for a $5$-form and the fact that $\overline\varepsilon_1
\gamma^{[3]}\varepsilon_1 = 0$ because of symmetry.  For the same
reason, the only contribution from the second term comes from the
contraction of $\gamma^n$ with $\gamma_{pq}$ in the spinor inner
product.  Using the trace identity \eqref{eq:trace} again, we arrive
at
\begin{equation*}
  \overline\varepsilon_1\gamma_m G^{(k)}\gamma_n\varepsilon_2\gamma^n
  \gamma^m \varepsilon_1 = -\half \overline\varepsilon_1\gamma_m
  \varepsilon_1 G^{(k)} \gamma^m \varepsilon_2 + \half
  \overline{\varepsilon}_1\gamma_m \varepsilon_1\gamma^{mn} G^{(k)}
  \gamma_n \varepsilon_2~.
\end{equation*}
Using the Clifford algebra and the trace identity \eqref{eq:trace}
once more we arrive at the desired identity.

\bibliographystyle{utphys}
\bibliography{AdS,AdS3,ESYM,Sugra,Geometry}

\end{document}